# Self consistent kinetic simulations of SPT and HEMP thrusters including the near-field plume region

**K. Matyash[1], R. Schneider, A. Mutzke, O. Kalentev**

*Max-Planck-Institut für Plasmaphysik, EURATOM Association, Greifswald, Germany*

**F. Taccogna**

*Istituto di Metodologie Inorganiche e di Plasmi IMIP-CNR, Sect. Bari, Bari, Italy*

**N. Koch and M. Schirra**

*THALES Electron Devices GmbH, Ulm, Germany*

**Abstract**

The Particle-in-Cell (PIC) method was used to study two different ion thruster concepts –Stationary Plasma Thrusters (SPT) and High Efficiency Multistage Plasma Thrusters (HEMP-T), in particular the plasma properties in the discharge chamber due to the different magnetic field configurations. Special attention was paid to the simulation of plasma particle fluxes on the thrusters' channel surfaces. In both cases, PIC proved itself as a powerful tool, delivering important insight into the basic physics of the different thruster concepts.

The simulations demonstrated that the new HEMP thruster concept allows for a high thermal efficiency due to both minimal energy dissipation and high acceleration efficiency. In the HEMP thruster the plasma contact to the wall is limited only to very small areas of the magnetic field cusps, which results in much smaller ion energy flux to the thruster channel surface as compared to SPT.

The erosion yields for dielectric discharge channel walls of SPT and HEMP thrusters were calculated with the binary collision code SDTrimSP. For SPT, an erosion rate on the level of 1 mm of sputtered material per hour was observed. For HEMP, thruster simulations have shown that there is no erosion inside the dielectric discharge channel.

## 1. Introduction

Ion thrusters are getting more and more important for scientific and commercial space missions, because they allow for a significant reduction of the spacecraft launch mass (by some 100 to 1000 kg). The reason for this is their by a factor of 5 to 10 increased specific impulse as compared to commonly used chemical thrusters [1]. As a consequence, commercial missions gain cost reductions and a larger flexibility in the choice of the launch rocket, whereas in case of scientific applications, missions deep into the solar system become possible. Also, fine tuning of the thrust correction can be done as needed e.g., for the compensation of atmospheric drag for low flying satellites. The necessary thrust for such applications ranges from micro-Newtons to some Newtons with electric input powers of some 10 to some 10000 Watts. The limits are set from the available power of the solar or nuclear electrical system on board of the satellite.

State-of-the-art ion thrusters for thrust levels above 0.1 mN ionize the propellant in a gas discharge and accelerate the produced ions by means of electrostatic fields to create a high energetic propulsive ion beam. Three main types of ion thruster technologies can be distinguished: grid ion thrusters (GITs), where propellant ionization and ion acceleration are separated by a system of grids, Hall Effect Thrusters HETs, of which Stationary Plasma Thrusters (SPTs) are a special case and High Efficiency Multistage Plasma Thrusters (HEMP-Ts). In case of HETs and HEMP-Ts, the plasma electron distribution is determined by magnetic fields such that propellant ionization and ion acceleration are self-consistently linked, however with different concepts for the magnetic field topology.

The SPT and HEMP thruster concepts result in quite different plasma-wall interaction characteristics. The SPT thruster relies on strong secondary electron emission from the dielectric walls of the thruster channel, which necessarily and self-consistently causes a large ion flux over the whole channel surface and consequently high erosion rate [2]. In contrast, in the HEMP thruster the plasma contact to the wall is limited only to very small areas of magnetic field cusps, which results in much smaller ion flux to the thruster channel surface as compared to SPT. Consequently, experimental studies of HEMP gave no evidence of erosion [3].

---
[1] e-mail: knm@ipp.mpg.de

This paper is dedicated to the numerical simulation of SPT and HEMP-T by means of the Particle-in-Cell (PIC) method. Particular emphasis is put on the differences of the plasma properties in the discharge chamber due to the different magnetic field configurations. We apply Particle-in-Cell code with Monte Carlo Collisions (PIC-MCC) to simulate the stationary operation regimes of both thrusters. The model resolves 2 spatial (rz) and 3 velocity components (2d3v). In the model all relevant collisional processes are included: electron-neutral elastic, ionization and excitation collisions, ion neutral momentum-transfer and charge exchange collisions. In the present simulations the computational domain is extended beyond the discharge channel and includes the near-field region of the thrusters.

The PIC MCC simulation delivers a full self-consistent microscopic description. Therefore, any plasma parameter of interest (both macroscopic and microscopic) can be calculated: potential and field, particle density, flux and temperature profiles, particle velocity distribution functions, etc. Thus, PIC simulations give us an important insight into the physics of the thruster, providing information which is difficult or impossible to get in experiments. The PIC method proved itself as a powerful tool suited for simulation of the large variety of low-temperature laboratory plasmas including ion thrusters plasmas [4], [5].

In chapter 2 the PIC MCC model and the simulation results for SPT are presented. Chapter 3 contains the simulation results for HEMP thruster and the discussion of the most important differences between the two thruster concept from the point of view of plasma confinement and plasma-surface interaction. Using ion flux distributions calculated with PIC MCC, the wall erosion of both thruster types is studied with the binary collision approximation (BCA) based Monte-Carlo code SDTrimSP [6]. Chapter 4 gives a summary comparing the different thruster features and discusses the outlook for future work.

## 2. SPT simulation

We apply a newly developed Particle-in-Cell code with Monte Carlo Collisions (PIC-MCC) to simulate the stationary operation regime of the SPT 100ML thruster [7]. SPT100 ML is the laboratory modification of the SPT100 thruster designed in the early 70s in the Kurchatov Institute, Moscow [8].

The detailed description of the PIC-MCC method can be found in thorough reviews [9], [10]. Here we just outline the main features of our model. In PIC-MCC simulation we follow the kinetics of so-called "Super Particles" (each of them representing many real particles), moving in the self consistent electric field calculated on a spatial grid from the Poisson equation. The particle collisions are handled by Monte-Carlo collision (MCC) routines, which randomly change particle velocities according to the actual collision dynamics. All relevant collisional processes are included in the model: electron-neutral elastic, ionization and excitation collisions, ion neutral momentum- transfer and charge exchange collisions. In order to self-consistently resolve anomalous electron transport due to near wall conductivity (NWC), a secondary electron emission (SEE) model for the thruster channel dielectric walls is included in the simulation. In our model we resolve 2 spatial (radial and axial) and 3 velocity components (2d3v). The computational domain in the present simulations is extended beyond the discharge channel and includes the near-field region of the thrusters.

The computational domain together with magnetic field topology is shown in Fig. 1. The computational domain represents a cylinder with length $Z_{max} = 80\,mm$ and radius $R_{max} = 80\,mm$. The thruster channel length (distance from the anode to the exit plane) is $Z_{thr} = 25\,mm$. The inner and outer thruster channel radii are $R_{in} = 34.5$ mm and $R_{out} = 50$ mm correspondingly. At $Z = 0$ the metal anode is located. The region outside the thruster exit: ($Z_{thr} < Z < Z_{max}$, $0 < R < R_{max}$) represents the near-field plume zone. The inner surface of the thruster channel and the exit plane are dielectric (Boron Nitride). The radial and axial boundaries of the computational domain are assumed to be metallic. All metal elements in the simulation, except for the anode are at the ground potential. At anode, the voltage $U_a = 300$ V is applied.

In order to compute the potential in the system, the Poisson equation:

$$\frac{1}{r}\varepsilon\frac{\partial\varphi}{\partial r}+\frac{\partial}{\partial r}\left(\varepsilon\frac{\partial\varphi}{\partial r}\right)+\frac{\partial}{\partial z}\varepsilon\frac{\partial\varphi}{\partial z}=-\frac{1}{\varepsilon_0}\rho \qquad (1)$$

is discretized on the grid, taking into account the possible change of the dielectric permittivity $\varepsilon$ across the dielectric surface. Applying the centered five-point scheme and assuming that at the point $(r_i, z_j)$ the dielectric surface can be both along the $R$ as well as along the $Z$ direction for the equidistant square grid ($\Delta r_i = \Delta z_j = \Delta r$) we get the finite difference equation for off-axis grid points ($r_i \neq 0$):

$$\begin{aligned}
&\left(\frac{1}{4r_i\Delta r}(\varepsilon_{i+1,j-1}+\varepsilon_{i+1,j+1})+\frac{1}{2\Delta r^2}(\varepsilon_{i+1,j-1}+\varepsilon_{i+1,j+1})\right)\varphi_{i+1,j}+ \\
&\left(\frac{1}{4r_i\Delta r}(\varepsilon_{i-1,j-1}+\varepsilon_{i-1,j+1}-\varepsilon_{i+1,j-1}-\varepsilon_{i+1,j+1})-\frac{1}{\Delta r^2}(\varepsilon_{i-1,j-1}+\varepsilon_{i-1,j+1}+\varepsilon_{i+1,j-1}+\varepsilon_{i+1,j+1})\right)\varphi_{i,j}+ \\
&\left(\frac{1}{2\Delta r^2}(\varepsilon_{i-1,j-1}+\varepsilon_{i-1,j+1})-\frac{1}{4r_i\Delta r}(\varepsilon_{i-1,j-1}+\varepsilon_{i-1,j+1})\right)\varphi_{i-1,j}+ \\
&\frac{1}{2\Delta r^2}(\varepsilon_{i+1,j+1}+\varepsilon_{i-1,j+1})\varphi_{i,j+1}+\frac{1}{2\Delta r^2}(\varepsilon_{i+1,j-1}+\varepsilon_{i-1,j-1})\varphi_{i,j-1}=-\frac{1}{\varepsilon_0}\cdot\frac{Q^V_{i,j}+Q^S_{i,j}}{V_{i,j}}
\end{aligned} \qquad (2)$$

Here $V_{i,j}$ is the grid cell volume, $Q^V_{i,j}$ and $Q^S_{i,j}$ are volume and surface charges associated with the cell. The dielectric permittivity is set to $\varepsilon_{i,j}=4$ inside the dielectric material (BN) and $\varepsilon_{i,j}=1$ inside the discharge channel and the near-field region.

For the points at the axis ($r_i=0$), taking into account that in the case of the axial symmetry $\left.\frac{\partial\varphi}{\partial r}\right|_{r=0}=0$ and $\lim_{r\to 0}\frac{1}{r}\varepsilon\frac{\partial\varphi}{\partial r}=\varepsilon\frac{\partial^2\varphi}{\partial r^2}$, (1) takes form:

$$2\varepsilon\frac{\partial^2\varphi}{\partial r^2}+\frac{\partial}{\partial z}\varepsilon\frac{\partial\varphi}{\partial z}=-\frac{1}{\varepsilon_0}\rho, \qquad (3)$$

and the corresponding finite difference equation is

$$\frac{1}{2\Delta r^2}(\varepsilon_{i+1,j-1}+\varepsilon_{i+1,j+1})\varphi_{i+1,j}-\frac{3}{\Delta r^2}(\varepsilon_{i+1,j-1}+\varepsilon_{i+1,j+1})\varphi_{i,j}+\frac{1}{\Delta r^2}\varepsilon_{i+1,j+1}\varphi_{i,j+1}+\frac{1}{\Delta r^2}\varepsilon_{i+1,j-1}\varphi_{i,j-1}=-\frac{1}{\varepsilon_0}\cdot\frac{Q^V_{i,j}+Q^S_{i,j}}{V_{i,j}}. \qquad (4)$$

If $(r_i, z_j)$ belongs to the metal element, then

$$\varphi_{i,j}=U_{i,j}. \qquad (5)$$

Here $U_{i,j}=300$ V for the anode and $U_{i,j}=0$ for the grounded metal parts. The resulting set of finite difference equations is solved using SuperLU decomposition library [11]. This approach allows us to calculate the potential inside the computational domain, self-consistently resolving the floating potential on the dielectric surfaces.

The magnetic field distribution used in the simulation (Fig. 1) corresponds to the Case 2 in [12] which is close to the one of the SPT100 ML thruster. We calculated the magnetic field using the freeware Finite-Element Magnetic Method solver (FEMM) [13]. The currents of magnetic coils and the magnetic cores and screens geometry were iteratively adjusted until the desired magnetic field topology was achieved. As reference the magnetic field lines topology and the dependence of the radial component of the magnetic field on the axial coordinate for Case 2 in [12] were used. The maximum radial magnetic field along the axial direction at the thruster median line between the channel walls in the simulation is $B_{max} = 180$ Gauss.

The simulation includes electrons, Xe$^+$ ions and the neutral Xenon atoms. Only charged particles dynamics is followed in the simulation. The particles equations of motion are solved for the discrete time steps with the leap-frog / Boris algorithm [9]. The neutral Xenon is treated as fixed background. The density of the neutral Xenon is taken from measurements reported in [14]. All surfaces in the simulation are assumed to be absorbing for plasma particles. At the dielectric surfaces of the thruster channel a simplified secondary electron emission model is applied: 10% of absorbed electrons are injected back starting from the same axial position with a kinetic energy corresponding to 90% of the incident energy and with a direction sampled randomly from a uniform distribution of the solid angle.

Electrons with a Maxwellian distribution and a temperature $T_e = 2$ eV are introduced into the system in the source regions at $34\,mm < Z < 36\,mm$ mm and $74\,mm < R < 76\,mm$. In the simulation the electrons from the source, accelerated in the thruster's electric field, are ionizing the neutral gas, creating the plasma in the thruster channel. In order to ensure an equilibrium solution, the electron source strength was adjusted during the simulation using a feed-back control loop.

To reduce the computational time the size of the system is scaled down by factor of 10. In order to preserve the ratio of the charged particles mean free paths and the gyroradii to the system length, the neutral Xenon density and the magnetic field are increased by the same factor 10.

An equidistant computational grid 160x160 was used in the simulation. A total of about 2000000 computational particles were used in the simulation. The cell size $\Delta R = \Delta Z = 5 \cdot 10^{-2}$ mm in the simulation was chosen to ensure that it is smaller than the smallest Debye length in the system. The time step was set to $\Delta t = 5.6 \cdot 10^{-12}$ s in order to resolve the electron plasma frequency. In total about ~ $10^7$ time steps were done before steady-state was reached.

The simulation was carried on a single processor of an Intel Xeon workstation. The duration of the run until an equilibrium solution was achieved was about 10 days.

The simulation results are summarized in Figs. 2-5. In Fig. 2 steady-state electron and Xe$^+$ densities profiles in the SPT100 ML thruster are presented. The electrons are accelerated from the neutralizer towards the anode, ionizing the neutral gas as they gain the energy from the electric field. The ion density closely follows the electron density, so that the resulting plasma potential profile is rather smooth, as one can see in Fig. 3. In our simulations the major potential drop of about 240 V takes place inside the thruster channel, whereas only 1/5 of the anode voltage falls after the exit plane. Thus, the exiting ions gain most of their energy inside the thruster channel. The plasma density reach its maximum in the discharge channel at about $Z \approx 12$ mm, where the electron impact ionization is most intense (see the ionization rate profile in Fig. 4). The electron temperature profile, shown in Fig. 5 indicates that the most efficient electron heating is occurring in the same region. The electrons get the energy from the electric field as they are gyrating in the magnetic field, which has a radial direction for most of the thruster channel, as they are trapped by the azimuthal *ExB* drift. As the mean free path for electron collisions for SPT operation parameters is much larger then the electron gyration radius, the efficiency of the electron heating rate has to be proportional to the energy electrons gain from the electric field on the Larmour radius. Thus, the most efficient heating should take place at the region where the ratio of the azimuthal electric field to the radial magnetic field $E_z/B_r$ has a maximum. In Fig. 6 we plot the ratio $E_z/B_r$ along the median line in the SPT100 ML channel. One can see, that the curve has its maximum at the axial position $Z \approx 12$ mm, where the maximum of the electron temperature is observed (see Fig. 5).

### 3. Simulation results for HEMP thruster and comparison with SPT

HEMP-Ts represent a new type of grid less ion thrusters with a particular magnetic confinement of the plasma electrons. The HEMP-T concept has been patented by Thales Electron Devices with an initial patent filed in 1998 [15]. A schematic view of the HEMP thruster concept with its magnetic field is shown in Fig. 7. In the HEMP-Ts the specific magnetic field topology provided by a sequential arrangement of magnetic stages with cusps efficiently confines the plasma electrons and minimizes plasma-wall contact. Electron movement towards the thruster anode is strongly impeded by this magnetic field topology to form steep electrical field gradients for effective ion acceleration. As a consequence, the HEMP thruster concept allows for a high thermal efficiency due to both minimal heat dissipation and high acceleration efficiency, and for a wide range of operational parameters. The latest HEMP thruster models have confirmed the following important features: broad and stable operational range in anode voltage and propellant mass flow and small heat dissipation to the thrusters.

We apply the same PIC MCC model, as described in the previous chapter, to simulate the HEMP thruster prototype DM3a [16]. The simulation computational domain together with calculated potential profile is shown in Fig. 8. The computational domain represents a cylinder with length $Z_{max} = 89\ mm$ and radius $R_{max} = 24\ mm$. The thruster length is $Z_{thr} = 51\ mm$. The thruster inner radius is $R_{thr} = 9\ mm$. At $Z = 0$ the metal anode is located. The major part of the thruster channel is dielectric. Only the outer ring at the exit is made of metal. The region outside the thruster exit: $Z_{thr} < Z < Z_{max}$ and $R < R_{max}$ represents the near-field plume zone. The radial and axial boundaries of the computational domain are assumed to be metallic. All metallic surfaces, except for the anode are at ground potential. At the anode, a voltage $U_a = 500$ V is applied. The electrons with a Maxwellian distribution and a temperature $T_e = 2$ eV are introduced into the system at the source region: $54\ mm < Z < 58\ mm$ mm and $16\ mm < R < 20\ mm$. To reduce the computational time the size of the system is scaled down by a factor of 10. An equidistant computational grid 890x240 was used in the simulation. A total of about 1000000 computational particles were used in the simulation. The cell size $\Delta R = \Delta Z = 10^{-2}$ mm and the time step $\Delta t = 1.12 \cdot 10^{-12}$ s were used in the simulation. In total about $\sim 10^6$ time steps were done before steady-state was reached.

Here, we present the most important simulation results in order to compare the two thrusters concepts (HEMP and SPT) in terms of their plasma confinement and plasma-surface interaction properties.

One of the main features of the HEMP thruster is the cusped magnetic field geometry, so that magnetic field lines are parallel to the side walls for most of the discharge channel and cross the surface only in small regions at the magnetic cusps. Such magnetic field geometry guarantees, that the plasma contact to the wall is limited only to the small areas of magnetic cusps as can be seen in the plasma density profiles obtained in the simulation in Fig. 9. This behavior is different from SPT where the magnetic field along the whole chamber is essentially radial, allowing plasma – wall contact along the whole channel as can be seen in Fig. 2. Actually, strong plasma-wall contact is a prerequisite for the SPT operation, as high electron fluxes to the side walls and resulting secondary electron emission are necessary for anomalous electron axial transport from the neutralizer to the anode due to near-wall conductivity. In HEMP the electrons can move freely along the magnetic filed lines between the cusps. In the cusps regions, where electrons are trapped in the azimuthal ExB drift, the diffusion due to electrons collisions with the neutral gas is enough to sustain the electron current toward the anode, necessary for the thruster operation. Thus, for the operation of the HEMP thruster the secondary electron emission from the channel walls is not necessary, as we also have seen in our simulations.

Another consequence of the magnetic field geometry for the HEMP thruster is that steep plasma potential drops are occurring only at cusps positions, whereas in the rest of the discharge the potential is rather flat (see the plasma potential profile in Fig. 8). The main potential drop, responsible for the ion beam acceleration, takes place at the exit cusp. Therefore, energetic ions in HEMP are generated only at the thruster exit. The uncompensated positive ion space charge outside the thruster is responsible for the steep potential drop at the exit, as the electrons are efficiently confined at the exit cusp by the radial magnetic field, as can be seen in Fig. 9. Thus, neutralization of the exiting ions space charge is not necessary for HEMP thruster operation, so that it can operate at much lower neutralizer currents as compared with SPT [17].

In the SPT, as one can see in Fig. 3 the ions are accelerated in the discharge channel starting from about 5 mm from the anode and reaching the energy ~ 140 eV close to the exit.

The differences in plasma confinement properties for two thruster concepts can be clearly seen in the plasma-wall fluxes profiles. In the left of Fig. 10 we show the ion fluxes to the dielectric discharge channel wall of the HEMP thruster and to the inner wall of SPT. One can see, as expected, that in the SPT the ion flux to the wall is continuous along the channel, whereas for the HEMP the ion flux is only non-zero at the anode cusp position, being below the model resolution of the dynamical range for particle fluxes everywhere else. The peak flux value at the cusp position in HEMP is higher than the maximum flux value in the SPT by factor 3. This is due to the higher operational plasma density for HEMP, as one can see in Figs. 2 and 9. However, as can be seen from the potential distribution inside HEMP, the energy of the ions hitting the wall at the anode cusp position should be at the level of 10-15 eV, which is well below the sputtering threshold ~50 eV for Boron Nitride ceramics under $Xe^+$ bombardment [18]. On the right of Fig. 10 we plot the mean energy of the ions hitting the wall along the channel for SPT and HEMP thruster calculated in the simulations. One can see that the energy of the ions impinging surface at the HEMP thruster anode cusp is below 15 eV, as it was expected from the potential profile. Thus, there should be no erosion of the dielectric discharge wall in the HEMP thruster. In the SPT, as it can be seen in Fig. 10 (right), the ions overcome the sputtering threshold of 50 eV already at $Z \approx 13$ mm, which can also be seen from the potential profile in Fig. 3, so one can expect wall erosion staring from this position.

In order to quantify the erosion yields for the dielectric walls in both thrusters, we performed the simulations with the binary collision SDTrimSP code [6].

SDTrimSP is a Monte Carlo program, which assumes an amorphous (randomized) target at zero temperature and infinite lateral size. The binary collision approximation is used to handle the atomic (nuclear) collisions. This means, that the change in flight direction due to the collision is given by the asymptotes of the real trajectory. For this evaluation the interaction potential is used to determine the scattering angle of the moving atom and recoil angle of the atom, which is set to motion. Then, the energy loss of the moving atom and the energy gain of recoils can be calculated. In addition, a moving atom transfers its energy to target electrons. The program follows projectiles (incident atoms) and target recoils atoms three-dimensionally until their energy falls below some preset value or if they have left the target (backscattering, transmission, sputtering). SDTrimSP can be used to calculate erosion yields, reflection coefficients, as well as more detailed information as depth distribution of implanted and energy distribution of backscattered and sputtered atoms.

We used the energy- angle- axial position- resolved ion fluxes calculated with PIC MCC code as input data for SDTrimSP simulations. The resulting erosion rates along the thruster wall for SPT100 ML and HEMP DM3a thrusters are presented in Fig 11. One can see that in the HEMP there is no erosion for the whole dielectric channel wall, including the anode cusp region. This supports the experimental observations of [3] where no erosion was found in the channel of HEMP thruster. The SDTrimSP results for SPT indicate erosion which starts as expected from $Z \approx 9$ mm and increases towards the exit, reaching 1.2 mm per 1000 hours. This agrees quite good with experimental investigation of SPT erosion in [2], where an erosion rate of about 5 mm per 1000 hours was reported, if one keeps in mind that our SPT100 ML simulations were performed for discharge current $I_d = 1.5$ A, which is a factor 3 lower than the discharge current measured in [2].

## 4. Conclusion

The Particle-in-Cell method was used to study two different ion thruster concepts – Hall Effect Thrusters and High Efficiency Multistage Plasma Thrusters - in particular the plasma properties in the discharge chamber due to the different magnetic field configurations. Special attention was paid to the simulation of the plasma particles fluxes onto the thruster channel surfaces. In both cases, PIC proved itself as a powerful tool, delivering important insight into the basic physics of the thrusters.

The simulations demonstrated that the new HEMP thruster concept allows for a high thermal efficiency due to both minimal energy dissipation and high acceleration efficiency. In the HEMP thruster the plasma contact to the wall is limited only to very small areas of the magnetic field cusps, which results in much smaller ion energy flux to the thruster channel surface as compared to SPT.

The erosion yields for dielectric discharge channel walls of SPT and HEMP thrusters were calculated with the binary collision code SDTrimSP. For the SPT the erosion rate on level of 1 mm of sputtered material per 1000 hours was observed. For HEMP thruster the simulations have shown that there is no erosion inside the dielectric discharge channel.

**Acknowledgments**
The authors want to thank the German Space Agency DLR for funding through Project No. 50 RS 0804. R.S. acknowledges funding of the work by the Initiative and Networking Fund of the Helmholtz Association.

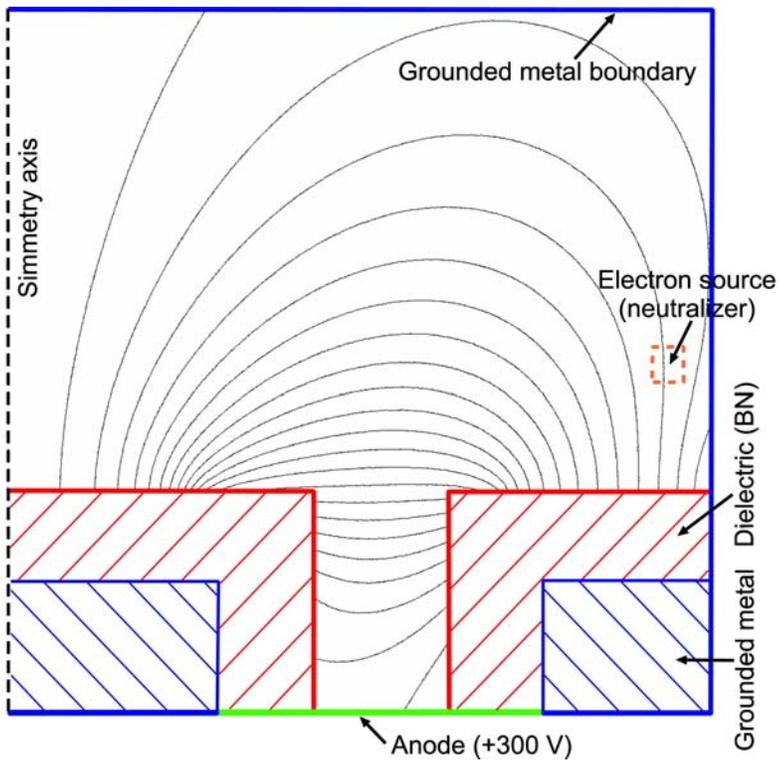

**Fig. 1.** Computational domain with magnetic field geometry for SPT100 ML simulations.

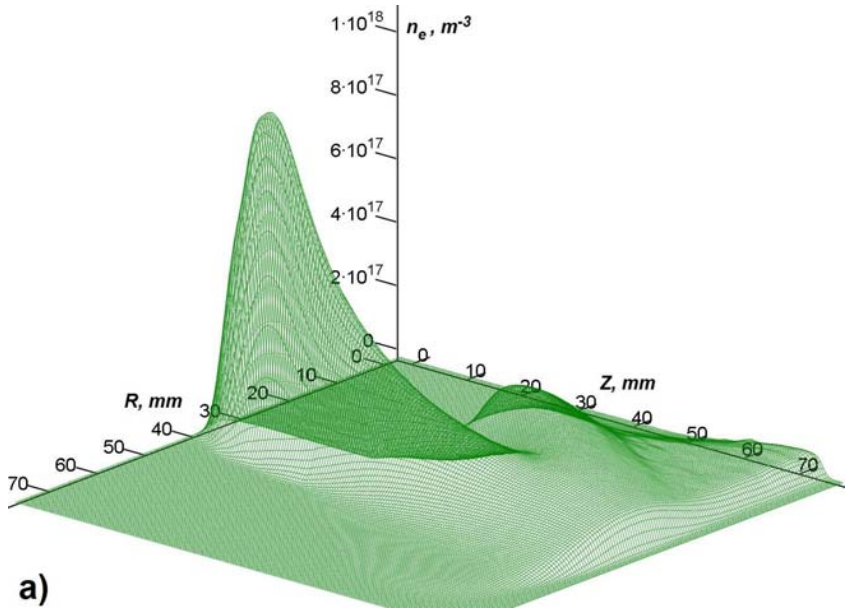

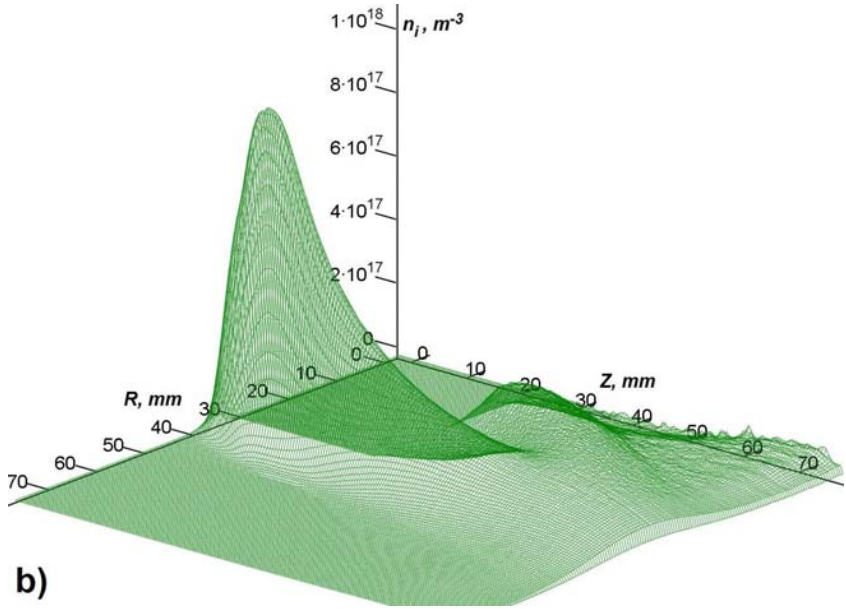

**Fig. 2.** Electron (a) and ion (b) density profiles in the SPT100 ML thruster.

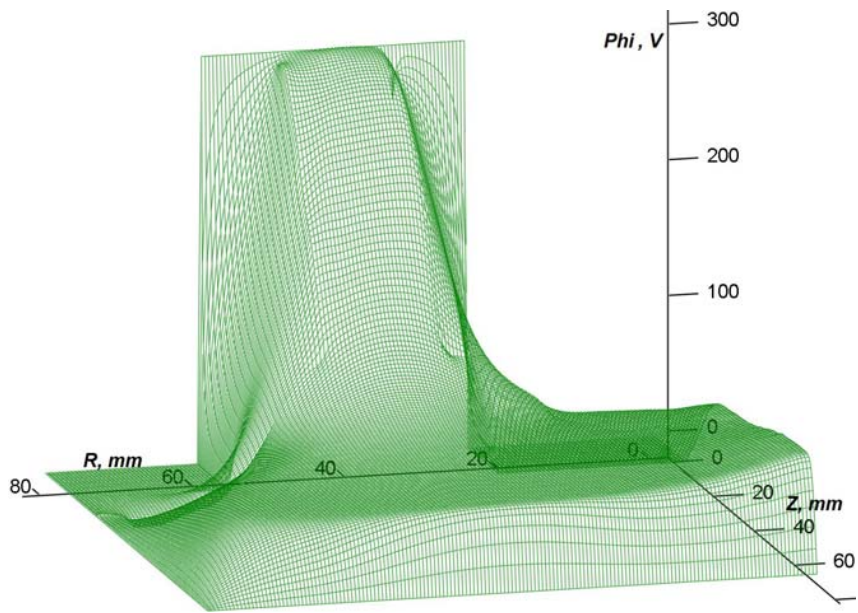

**Fig. 3.** Potential profile in the SPT100 ML thruster.

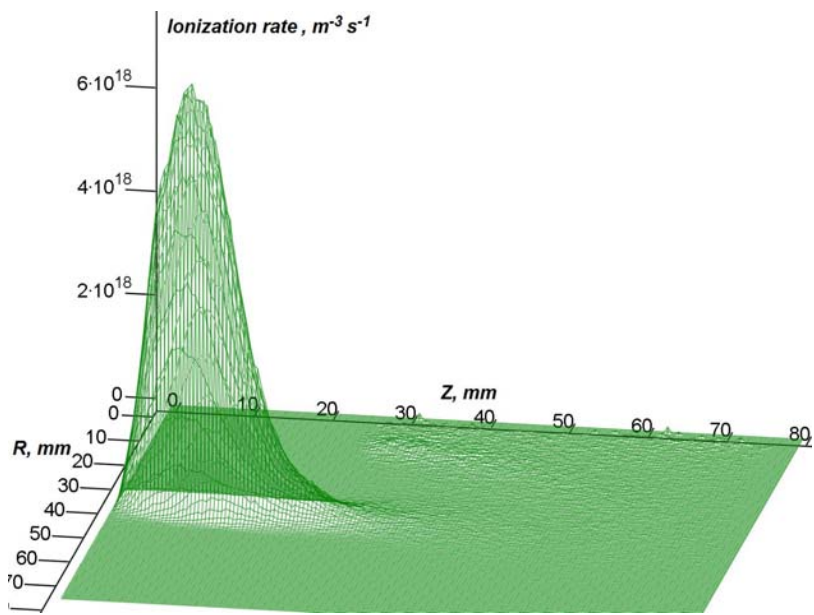

**Fig. 4.** Ionization rate profile in the SPT100 ML thruster.

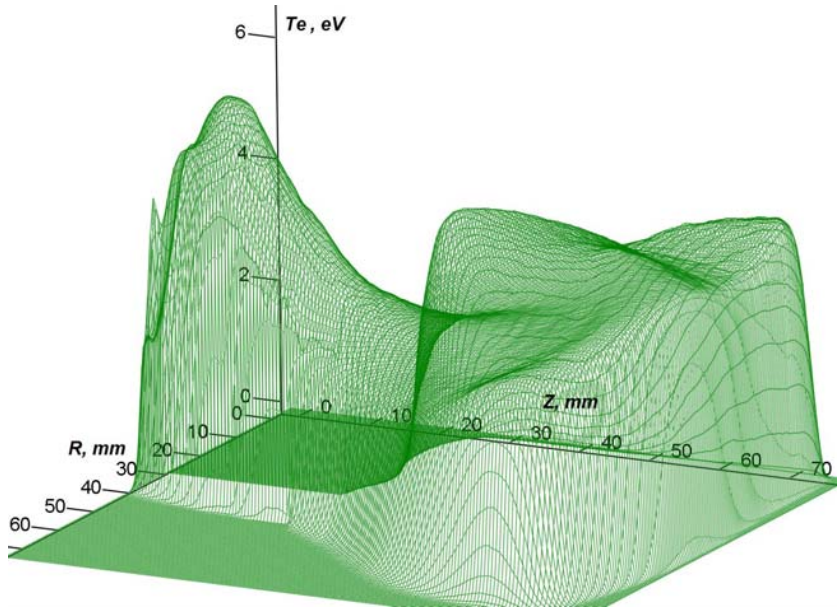

**Fig. 5.** Electron temperature profile in the SPT100 ML thruster.

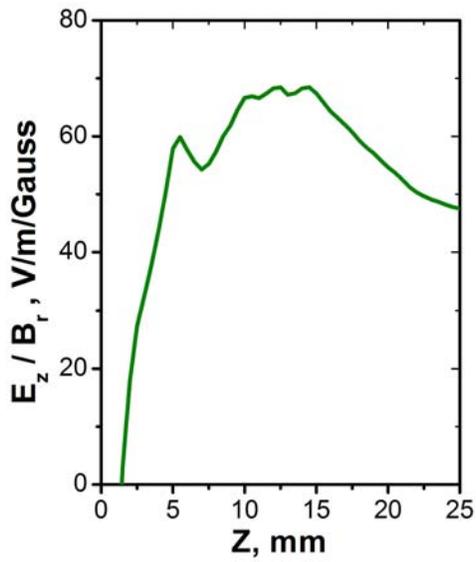

**Fig. 6.** Ez/Br ratio along the median line in the SPT100 ML channel.

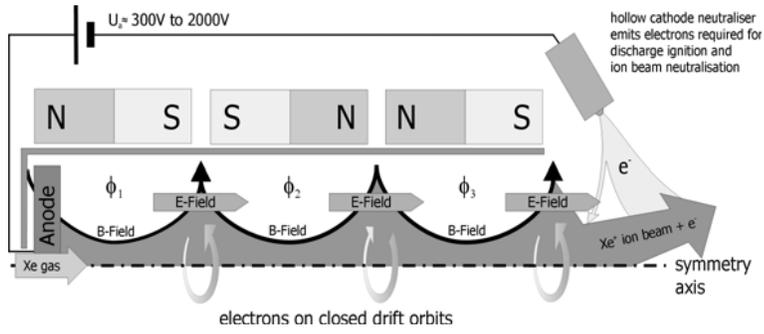

**Fig. 7.** Schematic view of the HEMP thruster concept. As an example, an arrangement with a cylindrical discharge channel and 3 magnet rings is shown, such that 3 cusp zones with high radial field are formed in the discharge channel.

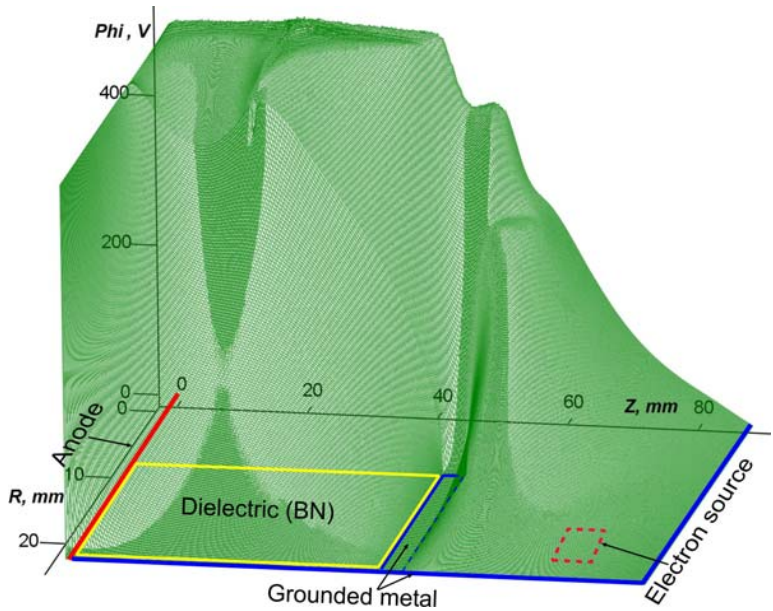

**Fig. 8.** Computational domain for the simulation of the HEMP DM3a thruster with a calculated potential profile.

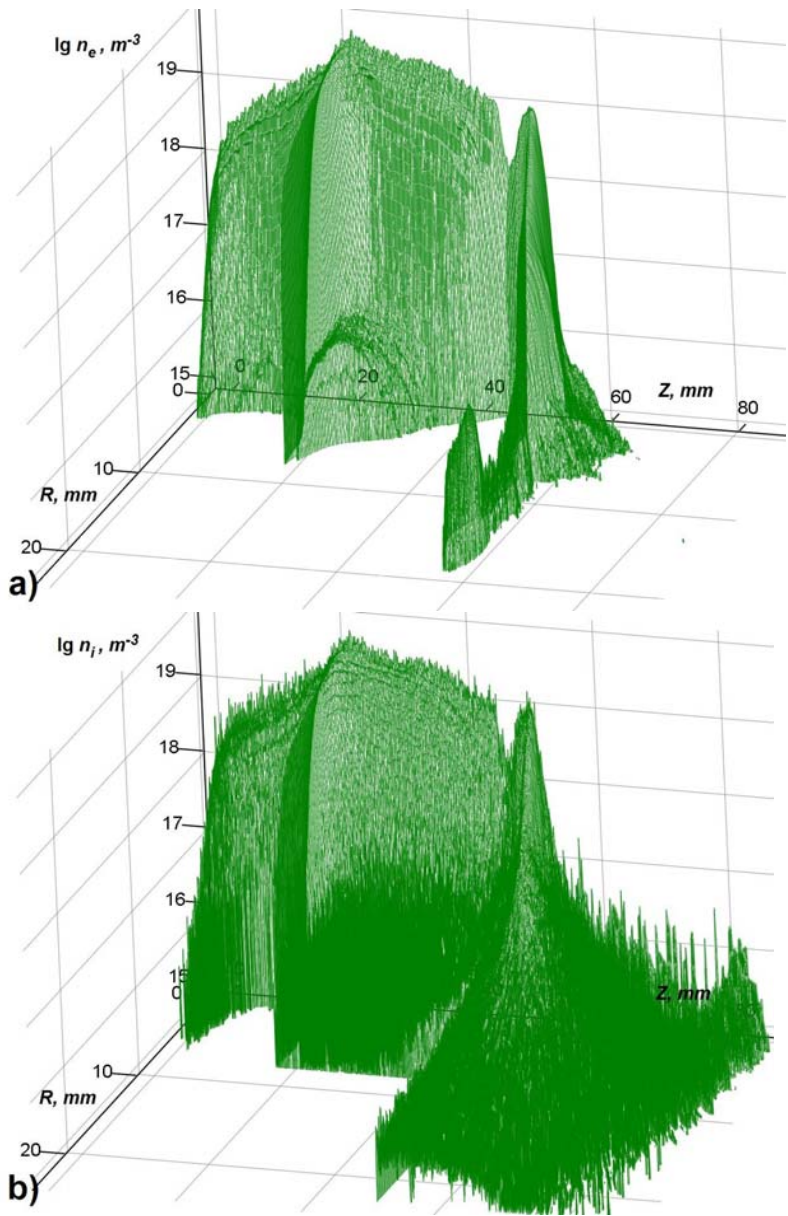

**Fig. 9.** Electron (a) and ion (b) density profiles in the HEMP DM3a thruster.

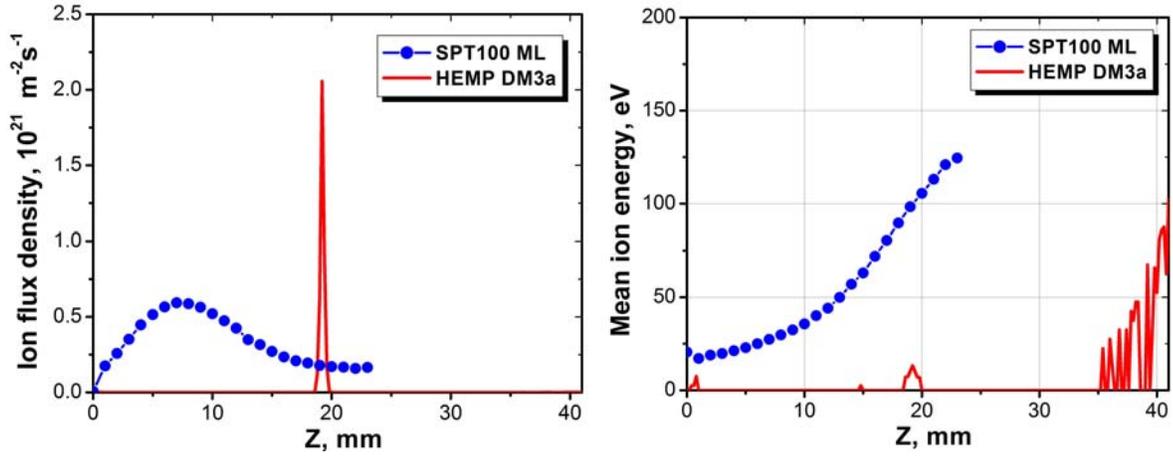

**Fig. 10.** The ion fluxes to the dielectric discharge channel wall of the HEMP DM3a thruster and to the inner wall of the SPT100 ML (left); corresponding mean ion energy (right).

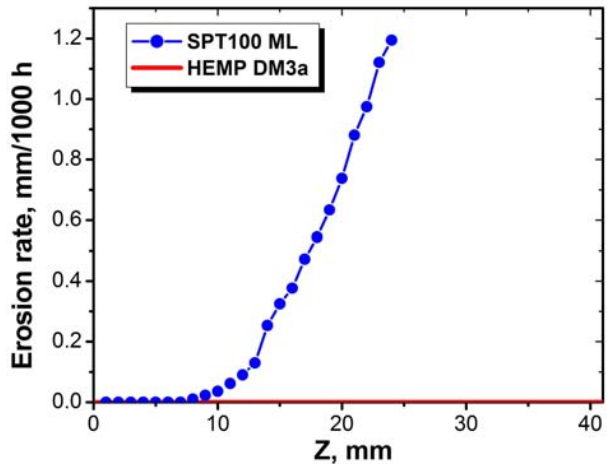

**Fig. 11.** The erosion rate for the dielectric discharge channel wall of the HEMP DM3a thruster and for the inner wall of SPT100 ML as calculated with SDTrimSP.